\documentclass{article}
\usepackage{spconf,amsmath,graphicx}
\usepackage{booktabs}
\usepackage{amsfonts}
\usepackage{color}
\usepackage[breaklinks=true]{hyperref}
\usepackage[anythingbreaks]{breakurl}
\usepackage{cite}  
\usepackage{multirow}  
\usepackage{footmisc}

\title{RECENT DEVELOPMENTS ON ESPNET TOOLKIT BOOSTED BY CONFORMER}
\name{
\begin{tabular}{c}
\it Pengcheng Guo$^{1,4}$, Florian Boyer$^{2,3}$, Xuankai Chang$^4$, Tomoki Hayashi${^{5}}$, Yosuke Higuchi$^6$\\
\it Hirofumi Inaguma${^7}$, Naoyuki Kamo$^8$, Chenda Li$^{9}$, Daniel Garcia-Romero$^4$, Jiatong Shi$^4$\\
\it Jing Shi$^{4,10}$, Shinji Watanabe$^4$, Kun Wei$^1$, Wangyou Zhang$^{9}$, Yuekai Zhang$^4$
\end{tabular}
}

\address{
    $^1$Northwestern Polytechnical University, $^2$LaBRI, University of Bordeaux, $^3$ Airudit \\
    $^4$Johns Hopkins University, $^5$Human Dataware Lab. Co., Ltd. \\
    $^6$Waseda University, $^7$Kyoto University, $^8$NTT Corporation $^{9}$Shanghai Jiao Tong University \\
    $^{10}$Institute of Automation, Chinese Academy of Sciences \\
}
\begin{document}
\ninept
\maketitle
\begin{abstract}
In this study, we present recent developments on ESPnet: End-to-End Speech Processing toolkit, which mainly involves a recently proposed architecture called Conformer, Convolution-augmented Transformer.
This paper shows the results for a wide range of end-to-end speech processing applications, such as automatic speech recognition (ASR), speech translations (ST), speech separation (SS) and text-to-speech (TTS).
Our experiments reveal various training tips and significant performance benefits obtained with the Conformer on different tasks.
These results are competitive or even outperform the current state-of-art Transformer models. 
We are preparing to release all-in-one recipes using open source and publicly available corpora for all the above tasks with pre-trained models.
Our aim for this work is to contribute to our research community by reducing the burden of preparing state-of-the-art research environments usually requiring high resources.

\end{abstract}
\section{Introduction}
\label{sec:intro}
Transformer architecture has drawn immense interest recently and became the dominated model due to its effectiveness across various sequence-to-sequence tasks, like machine translation,  language modeling (LM), and automatic speech recognition (ASR)~\cite{vaswani2017attention,devlin2018bert,dai2019transformer,dong2018speech,zeyer2019comparison,karita2019comparative}. One reason for the success of Transformer model is that the multihead self-attention layers can learn long-range global context better than the recurrent neural networks (RNNs). However, for speech processing tasks, not only the global context, but also the local information is crucial to capture some particular properties of speech, like coarticulation and monotonicity. Convolution neural networks (CNNs), on the other hand, are good at extracting fine-grained local feature patterns. Recently, Gulati \textit{et al.} ~\cite{gulati2020conformer}~proposed a novel architecture with combination of self-attention and convolution in ASR models, which is named \textit{Conformer}. With this proposed design, the self-attention layer learns the global context while the convolution module efficiently captures the local correlations synchronously.

In addition to the ASR task, other speech processing tasks can also have such benefits and obtain improvement when giving local information. In this study, we aim to explore the efficiency of Conformer on various end-to-end speech processing applications, including automatic speech recognition (ASR), speech translation (ST), speech separation (SS), and text-to-speech (TTS). We provide intensive comparisons of Conformer with Transformer on lots of publicly available corpora and try our best to share the practical guides (e.g., learning rate, hyper-parameters, network structure) on the use of Conformer. We also prepare to release the reproducible recipes and state-of-the-art setups to the community to succeed our exciting outcomes.

The contributions of this study include:
\begin{itemize}
    \item We extend the Conformer architecture to various end-to-end speech processing applications and conduct comparative experiments with Transformer.
    \item We share our practical guides for the training of Conformer, like learning rate, kernel size of Conformer block, and model architectures, etc. 
    \item We provide reproducible benchmark results, recipes, setups and well-trained models on a large number of publicly available corpora\footnote{Due to the page limitation, we are not able to cite all references. Instead, corresponding links are embed in the corpora names.} in our open source toolkit ESPnet~\cite{watanabe2018espnet,inaguma2020espnet,hayashi2020espnet}.
\end{itemize}


\section{CONFORMER}
\label{sec:conformer}

Our Conformer model consists of a Conformer encoder proposed in~\cite{gulati2020conformer} and a Transformer decoder. The encoder is a multi-blocked architecture and each block is stacked by a positionwise feed-forward (FFN) module, a multihead self-attention (MHSA) module, a convolution (CONV) module, and another FFN module in the end. We apply layer normalization (LN) before each module and dropout followed by a residual connection afterward (pre-norm), as in~\cite{zeyer2019comparison,wang2019learning}. This section describes the details of each module in the encoder.

\subsection{Multihead self-attention module}
\label{subsec:selfatt}
The idea of MHSA module is to learn an alignment in which each token in the sequence learns to gather from other tokens~\cite{bahdanau2014neural,tay2020efficient}. For each single head $h$, the output of attention computation can be formulated as:
\begin{equation}
    \text{Att}(\mathbf{Q}_h, \mathbf{K}_h, \mathbf{V}_h) = \text{Softmax} \left( \frac{\mathbf{Q}_h \mathbf{K}^T_h}{\sqrt{d^{\text{att}}}} \right) \mathbf{V}_h,
\end{equation}
where $\mathbf{Q}_h = \mathbf{W}_h^q \mathbf{X}$, $\mathbf{K}_h = \mathbf{W}_h^k \mathbf{X}$, and $\mathbf{V}_h = \mathbf{V}_h^q \mathbf{X}$ are query, key, and value linear transformations applied on the input sequence $\mathbf{X} \in \mathbb{R}^{T \times d^{\text{att}} }$. $\mathbf{W}_h^q, \mathbf{W}_h^k, \mathbf{W}_h^v \in \mathbb{R}^{d^{\text{att}} \times \frac{d^{\text{att} } }{H}}$ are the projecting weight matrices, $d^{\text{att}}$ is the dimension of attention, and $H$ refers to the total number of attention heads. The term $\frac{1}{\sqrt{d^{\text{att}}}}$ is used to scale the dot product result to avoid a very large magnitude caused by the dimension of attention. In order to jointly attend to information from different representation subspaces, the outputs of each head are concatenated together and fed into a fully-connected layer, as follows:
\begin{align}
    \text{MHSA}(\mathbf{Q}, \mathbf{K}, \mathbf{V}) &= \text{Concat}(\text{head}_1,...,\text{head}_{H})\mathbf{W}^o, \\
    \text{head}_h &= \text{Att}(\mathbf{Q}_h, \mathbf{K}_h, \mathbf{V}_h),
\end{align}
where $\mathbf{W}^o \in \mathbb{R}^{d^{\text{att}} \times d^{\text{att}}}$ is an output linear projecting matrix.

Besides, Conformer also integrates a position encoding scheme from Transformer\-XL~\cite{dai2019transformer} to generate better position information for the input sequence with various lengths, named relative positional encodings. For an input sequence $\mathbf{X}$, the computational procedure can be summarized as:
\begin{equation}
    \mathbf{X} = \mathbf{X} + \text{Dropout}( \text{MHSA}( \text{LN}(\mathbf{X}) ) ).
\end{equation}

\subsection{Convolution module}
\begin{figure}[htbp]
    \centering
    \includegraphics[width=0.8\linewidth]{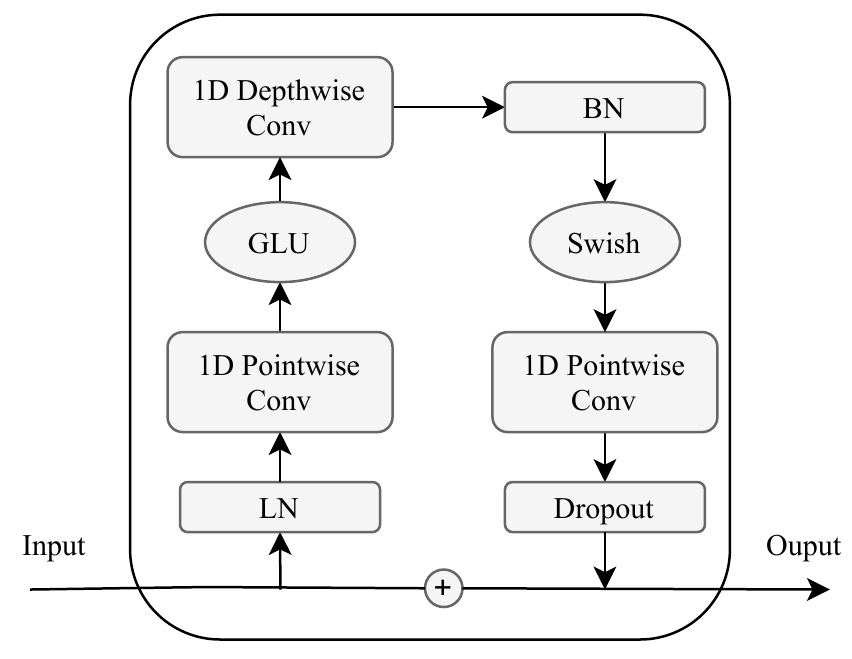}
    \caption{Details of the CONV module. All convolution operations are performed on the time domain.}
    \label{fig:cnn-layer}
    \vspace{-0.2cm}
\end{figure}

Figure~\ref{fig:cnn-layer} illustrates the details of CONV module. The CONV module starts with a 1-D pointwise convolution layer and a gated linear units (GLU) activation~\cite{dauphin2017language}. The 1-D pointwise convolution layer doubles the input channels, while the GLU activation splits the input along the channel dimension and conducts an element-wise product. After that, it is followed by a 1-D depthwise convolution layer, a batch normalization (BN) layer, a Swish activation, and another 1-D pointwise convolution layer. 

\subsection{Pointwise feed-forward module}
The FFN module in original Transformer is composed of two linear transformations with a ReLU activation in between, as follows:
\begin{equation}
    \text{FFN}(\textbf{X}) = \mathbf{W}_2 \text{ReLU}( \mathbf{W}_1\mathbf{X} + b_1) + b_2,
\end{equation}
where $\mathbf{W}_1 \in \mathbb{R}^{d^{\text{att}} \times d^{\text{ff}}}$, $\mathbf{W}_2 \in \mathbb{R}^{d^{\text{ff}} \times d^{\text{att}}}$ are linear projecting matrix and $d^{\text{att}}$ denotes the hidden dimension of linear layer.

Different from Transformer, Conformer introduces another FFN module and replaces the ReLU activation with the Swish activation. Besides, inspired by Macaron-Net~\cite{lu2019understanding}, the two FFN modules are following a half-step scheme and sandwiching the MHSA and CONV modules. Mathematically, for input $\mathbf{X}$, the output is:
\begin{equation}
	\mathbf{X} = \mathbf{X} + \frac{1}{2} \times \text{Dropout} ( \text{FFN}( \text{LN}(\mathbf{X}) ) ).
\end{equation}

\subsection{Conformer block}
\begin{figure}[tb]
    \centering
    \includegraphics[width=0.95\linewidth]{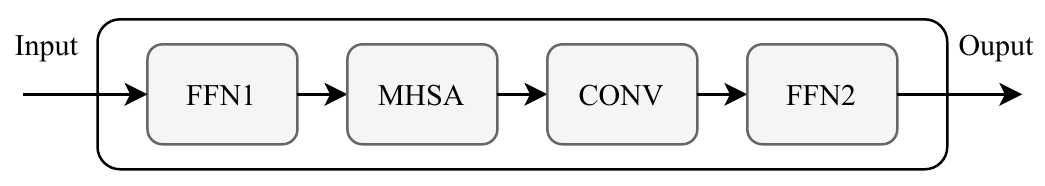}
    \caption{An overview of Conformer block.}
    \label{fig:conformer}
    \vspace{-0.2cm}
\end{figure}

Figure~\ref{fig:conformer} shows how to combine each module together. The difference between the Conformer block and Transformer block include: the relative positional encoding, the integrated CONV module, and a pair of FNN modules in the Macaron-Net style.

\section{SPEECH APPLICATIONS}
\label{sec:speech-task}
In ASR tasks, the Conformer model predicts a target sequence $Y$ of characters or byte-pair-encoding (BPE) tokens\footnote{SentencePiece toolkit~\cite{kudo2018sentencepiece} is used to generate the BPE tokens.} from an input sequence $\mathbf{X}$ of 80 dimensional log-mel filterbank features with/without 3-dimensional pitch features. 
$\mathbf{X}$ is first sub-sampled in a convolutional layer by a factor of 4, as in~\cite{dong2018speech}, and then fed into the encoder and decoder to compute the cross-entropy (CE) loss. The encoder output is also used to compute a connectionist temporal classification (CTC) loss~\cite{graves2006connectionist} for joint CTC-attention training and decoding~\cite{kim2017joint}. During inference, token-level or word-level language model (LM)~\cite{hori2018end} is combined via shallow fusion.

ST tasks adopt the same framework defined in ASR. It directly maps speech from a source language to the corresponding translation in the target language. In order to eliminate the serious under-fitting problem, we initialize the ST encoder by a pre-trained ASR encoder and start the ST decoder from a pre-trained machine translation (MT) decoder, as in~\cite{inaguma2020espnet}.

For the SS tasks, the Conformer model is optimized to estimate the time-frequency mask for each individual speaker given a speech mixture. The model is trained with utterance-level permutation invariant loss (uPIT)~\cite{kolbaek2017multitalker}. Different from the ASR system, the Conformer model here only contains the encoder, followed by an additional linear layer and activation function to predict the masks.

TTS tasks use the Conformer encoder for non-autoregressive TTS models~\cite{ren2019fastspeech,ren2020fastspeech,lancucki2020fastpitch}, which generates a sequence of log-mel filterbank features from a phoneme or character sequence in cooperation with the duration predictor~\cite{ren2019fastspeech}. The whole model is optimized to minimize the L1 loss for the target features and the mean square error (MSE) loss for the durations.

\begin{table*}[htbp]
\centering
\caption{CER/WER results on various open source ASR corpora. Both Transformer and Conformer models are implemented based on ESPnet toolkit. $\ast$ marks ESPnet2 results. $\dagger$ and $\ddagger$ indicate only w/ speed or  only w/ SpecAugment, respectively. $\S$ denotes w/o any data augmentation.}
\vspace{0.2cm}
\label{table:asr-results}
\resizebox{1.0\linewidth}{!}{
    \begin{tabular}{ccc|c|c|c} 
    \toprule[1.5pt]
    Dataset & Vocab & \multicolumn{1}{c}{Metric} & \multicolumn{1}{c}{Evaluation Sets} & \multicolumn{1}{c}{Transformer} & Conformer \\ 
    \hline\hline
    \href{https://openslr.magicdatatech.com/62/}{AIDATATANG} & Char & CER & dev / test & ($\dagger$) 5.9 / 6.7 & \textbf{4.3} / \textbf{5.0} \\ 
    \href{https://openslr.magicdatatech.com/33/}{AISHELL-1} & Char & CER & dev / test & ($\dagger$) 6.0 / 6.7  & (*) \textbf{4.4} / \textbf{4.7} \\
    \href{http://www.aishelltech.com/aishell_2}{AISHELL-2} & Char & CER & android / ios / mic & ($\dagger$) 8.9 / 7.5 / 8.6 & \textbf{7.6} / \textbf{6.8} / \textbf{7.4} \\
    \href{http://aurora.hsnr.de/aurora-4.html}{AURORA4} & Char & WER & dev\_0330 (A / B / C / D) & \textbf{3.3} / \textbf{6.0} / \textbf{4.5} / 10.6 & 4.3 / 6.0 / 5.4 / \textbf{9.3} \\
    \href{https://pj.ninjal.ac.jp/corpus_center/csj/en/}{CSJ} & Char & CER  & eval\{1, 2, 3\} & ($\ast$) 4.7 / 3.7 / 3.9 & ($\ast$) \textbf{4.5 } / \textbf{3.3} / \textbf{3.6}  \\
    \href{http://spandh.dcs.shef.ac.uk/chime_challenge/chime2016/}{CHiME4} & Char & WER & \{dt05, et05\}\_\{simu, real\} & ($\dagger$) 9.6 / 8.2 / 15.7 / 14.5 & \textbf{9.1} / \textbf{7.9} / \textbf{14.2} / \textbf{13.4} \\
    \href{https://catalog.ldc.upenn.edu/LDC2014T23}{Fisher-CallHome} & BPE & WER & dev / dev2 / test / devtest / evltest & 22.1 / 21.5 / 19.9 / 38.1 / 38.2 & \textbf{21.5} / \textbf{21.1} / \textbf{19.4} / \textbf{37.4} / \textbf{37.5}  \\
    \href{https://catalog.ldc.upenn.edu/LDC2005S15}{HKUST} & Char & CER & dev & ($\dagger$) 23.5 & ($\dagger$) \textbf{22.2} \\
    \href{https://sites.google.com/site/shinnosuketakamichi/publication/jsut}{JSUT} & Char & CER & our split & ($\dagger$) 18.7 & \textbf{14.5} \\
    \href{https://openslr.magicdatatech.com/12/}{LibriSpeech} & BPE & WER & \{dev, test\}\_\{clean, other\} & 2.1 / 5.3 / 2.5 / 5.5 & \textbf{1.9} / \textbf{4.9} / \textbf{2.1} / \textbf{4.9} \\
    \href{https://reverb2014.dereverberation.com}{REVERB} & Char & WER & et\_near / et\_far & ($\dagger$) 13.1 / 15.4 & ($\dagger$) \textbf{10.5} / \textbf{13.9} \\
    \href{https://catalog.ldc.upenn.edu/LDC97S62}{Switchboard} & BPE & WER & eval2000 (callhm/ swbd) & 17.3 / 8.5 & \textbf{15.0} / \textbf{7.1} \\
    \href{https://openslr.magicdatatech.com/19/}{TEDLIUM2} & BPE & WER  & dev / test & 9.3 / 8.1 & \textbf{8.6} / \textbf{7.2} \\
    \href{https://openslr.magicdatatech.com/51/}{TEDLIUM3} & BPE & WER & dev / test & 10.8 / 8.4 & \textbf{9.6} / \textbf{7.6} \\
    \href{https://openslr.magicdatatech.com/49/}{VoxForge} & Char & CER & our split & ($\S$) 9.4 / 9.1 & ($\S$) \textbf{8.7} / \textbf{8.2} \\
    \href{https://catalog.ldc.upenn.edu/LDC94S13A}{WSJ} & BPE & WER & dev93/ eval92 & ($\ddagger$) \textbf{7.4} / \textbf{4.9} & ($\ddagger$) 7.7 / 5.3  \\
    WSJ-2mix & Char & WER & tt & ($\S$) 12.6 & ($\S$) \textbf{11.9} \\
    \bottomrule[1.5pt]
    \end{tabular}
}
\vspace{-0.4cm}
\end{table*}

\section{SPEECH RECOGNITION EXPERIMENTS}
\label{sec:asr-exp}

\subsection{Setups}
\label{subsec:asr-setup}
To evaluate the effectiveness of our Conformer model, we conduct experiments on a total of 25 ASR corpora, including various recording environments (clean, noisy, far-field, mixed speech), languages (English, Mandarin, Japanese, Spanish, low-resource languages), and sizes (10 - 960 hours). Most of the corpora are followed the same data preparation procedure as in Kaldi~\cite{povey2011kaldi}. Optionally, we also use speed perturbation~\cite{ko2015audio} at ratio 0.9, 1.0, 1.1 and SpecAugment~\cite{park2019specaugment} for the data augmentation in some corpora.

For each corpus, the detail configurations of our Conformer model are same as ESPnet Transformer recipes~\cite{karita2019improving} ($\text{Enc}=12, \text{Dec}=6, d^{\text{ff}}=2048, H=4, d^{\text{att}}=256$). Particularly, the number of attention heads and attention dimensions are different for Librispeech ($H=8, d^{\text{att}}=512$). The convolution subsampling layer has 2-layer CNN with 256 channels, stride with 2, and kernel size with 3. For different corpora, we train 20-100 epochs and average the last 10 best checkpoints as the final model. We tune the learning rate coefficient (e.g., 1-10) and the kernel size of CONV module (e.g., 5-31) on the corresponding development sets to obtain the best results. Detail setups can be referred to ESPnet recipes\footnote{\url{https://github.com/espnet/espnet}\label{foot:espnet}}.

\begin{table}[tbp]
\vspace{-0.2cm}
\centering
\caption{WER results on dev/test sets of low-resource language corpora. BPE tokens are used as the output units.}
\vspace{0.2cm}
\label{table:asr-lowresource-results}
\resizebox{0.85\linewidth}{!}{
    \begin{tabular}{c|c|c} 
    \toprule[1.5pt]
    \multicolumn{1}{c}{Dataset} & \multicolumn{1}{c}{Transformer} & \begin{tabular}[c]{@{}c@{}}Conformer\\+ Data Augmentation\end{tabular} \\ 
    \hline\hline
    \href{https://www.openslr.org/89/}{Yolox\'ochitl-Mixtec} & 23.0 / 23.2 & \textbf{16.0} / \textbf{16.1} \\
    \href{https://www.openslr.org/92}{Puebla-Nahuat} & 27.9 / 26.0 & \textbf{23.5} / \textbf{21.7} \\
    \href{https://commonvoice.mozilla.org/en/datasets}{Commonvoice-Czech} & 38.2 / 44.3 & \textbf{15.3} / \textbf{20.6} \\
    \href{https://commonvoice.mozilla.org/en/datasets}{Commonvoice-Welsh} & 32.0 / 21.8 & \textbf{20.0} / \textbf{14.2} \\
    \href{https://commonvoice.mozilla.org/en/datasets}{Commonvoice-Russian} & 22.0 / 27.3 & \textbf{6.9} / \textbf{8.5} \\
    \href{https://commonvoice.mozilla.org/en/datasets}{Commonvoice-Italian} & 31.8 / 33.7 & \textbf{15.6} / \textbf{17.0} \\
    \href{https://commonvoice.mozilla.org/en/datasets}{Commonvoice-Persian} & 8.5 / 10.2  &  \textbf{1.4} / \textbf{2.1} \\
    \href{https://commonvoice.mozilla.org/en/datasets}{Commonvoice-Polish} & 24.1 / 15.1 & \textbf{8.8} / \textbf{2.6} \\
    \bottomrule[1.5pt]
	\end{tabular}
}
\vspace{-0.2cm}
\end{table}
\vspace{-0.2cm}

\subsection{Results}
\label{subsec:asr-result}
Table~\ref{table:asr-results} shows the character and word error rate (CER/WER) results on each corpus. It can be seen that Conformer model outperforms Transformer on 14/17 corpora in our experiments and even achieves state-of-the-art results on several corpora, like AIDATATANG and AISHELL-1. Instead of the single-speaker speech, it also brings about 7\% relative improvement compared with Transformer on the multi-speaker WSJ-2mix data. Besides, we also conduct experiments to investigate the generalization of Conformer models on low-resource language corpora, as shown in Table~\ref{table:asr-lowresource-results}. Conformer achieves more than 15\% relative improvements in all 8 different languages compared with Transformer model. 

Since our Conformer model uses the same decoder framework as Transformer, the performance gains may come from the additional local information provided by the CONV module. Thus, we study the effects of the CONV module by training a pure CTC model or a Transducer model with the Conformer encoder. Table~\ref{table:asr-ctc-results} summaries the CER/WER results of two pure CTC models, while Table~\ref{table:asr-rnnt-results} shows the CER results of different Transducer models. We use a single-LSTM layer decoder in all Transducer models. Detail setups can be referred to ESPnet recipes.\footref{foot:espnet}. Both Conformer-CTC and TDNN-Conformer-Transducer models show consistent improvement and the Conformer-CTC model even achieves competitive results over Transformer with a decoder. From above results, we can conclude that Conformer shows superior performance in various types of ASR corpora, even in the challenging far-field, mixed speech, and low-resource language scenarios.

\begin{table}[tbp]
\vspace{-0.2cm}
\centering
\caption{CER/WER results of pure CTC models. Vocabularies, error metrics and evaluations sets are same as Table~\ref{table:asr-results}. We only use greedy-search without LM rescoring during inference.}
\vspace{0.2cm}
\label{table:asr-ctc-results}
\resizebox{0.8\linewidth}{!}{
    \begin{tabular}{c|c|c} 
    \toprule[1.5pt]
    \multicolumn{1}{c}{Dataset} & \multicolumn{1}{c}{Transformer-CTC} & Conformer-CTC \\ 
    \hline\hline
    \href{https://pj.ninjal.ac.jp/corpus_center/csj/en/}{CSJ} & 6.0 / 4.2 / 4.8 & \textbf{4.8} / \textbf{3.7} / \textbf{3.8} \\
    \href{https://openslr.magicdatatech.com/19/}{TEDLIUM2} & 16.7 / 16.6 & \textbf{9.3} / \textbf{8.7} \\
    \href{https://openslr.magicdatatech.com/49/}{VoxForge} & 14.0 / 14.1 & \textbf{9.2} / \textbf{8.4} \\
    \href{https://catalog.ldc.upenn.edu/LDC94S13A}{WSJ} & 19.4 / 15.5 & \textbf{12.9} / \textbf{10.9} \\
    \bottomrule[1.5pt]
	\end{tabular}
}
\vspace{-0.2cm}
\end{table}


\begin{table}[tbp]
\vspace{-0.2cm}
\centering
\caption{CER results of different Transducer models on the dev/test set of the \href{https://ailab.hcmus.edu.vn/vivos/}{VIVOS} corpus. All experiments w/o data augmentation.}
\vspace{0.2cm}
\label{table:asr-rnnt-results}
\resizebox{0.7\linewidth}{!}{
    \begin{tabular}{c|c|c} 
    \toprule[1.5pt]
    \multicolumn{1}{c}{Model} & \multicolumn{1}{c}{dev} & test \\ 
    \hline\hline
    Transformer-Transducer & 17.2 & 17.1 \\
    Conformer-Transducer & 13.7 & 14.0 \\
    TDNN-Conformer-Transducer & \textbf{11.6} & \textbf{13.1}  \\
    \bottomrule[1.5pt]
	\end{tabular}
}
\vspace{-0.2cm}
\end{table}

\vspace{-0.2cm}
\subsection{Discussion}
\label{asr-dis}
Following are some training tips from our experiments:
\begin{itemize}
	\item When Conformer occurs a sudden accuracy drop on the training set, decreasing the learning rate can lead to more stable training. We use the learning rate in \{1, 2, 5, 10\} for different corpora.
	\item The kernel size of the CONV module is related to the input sentence length in the corpora. We use the kernel size in \{5, 7, 15, 31\} for different corpora.
	\item In addition to the warmup training strategy~\cite{vaswani2017attention}, the OneCycleLR~\cite{smith2019super} learning scheduler can also give a stable training of self-attention based models.
\end{itemize}


\section{SPEECH TRANSLATION EXPERIMENTS}\label{st-exp}
The configuration of all our ST models are same as ASR systems described in Sec~\ref{subsec:asr-setup}. During the training, we initialized the model parameters with the pre-trained encoder and decoder optimized on ASR and MT parallel data involved in each corpus, respectively. We conduct the ST experiment on the \href{https://catalog.ldc.upenn.edu/LDC2014T23}{Fisher-CallHome Spanish} corpus, and evaluate on five common test sets. We use the \texttt{Fisher-dev} set as the development set. The input speech feature is same as the ASR system and the output tokens are 1k BPE tokens. Same data augmentation techniques are used to improve the performance. 

The Conformer model achieves about 10\% relative improvement over the baseline Transformer model in the ST task as well. To validate the gains did not come from just increasing the model parameters with additional CONV and FFN modules, we also train a Conformer-small model by decreasing $d^{\text{ff}}$ from 2048 to 1024 to keep the parameter budget for a fair comparison. Although the BLEU score is slightly decreased by halving $d^{\text{ff}}$, our Conformer-small model still significantly outperforms the Transformer model.


\begin{table}[htbp]
\centering
\caption{BLEU scores on Fisher-CallHome Spanish corpus. 1k SentencePiece tokens are used as the output units.}
\vspace{0.3cm}
\label{table:st-results}
\resizebox{1.\linewidth}{!}{
    \begin{tabular}{c|c|c|c|c|c|c} 
    \toprule[1.5pt]
    \multirow{2}{*}{Model} & \multirow{2}{*}{\text{\#Param [M]}} & \multicolumn{3}{c|}{Fisher} & \multicolumn{2}{c}{CallHome} \\ \cline{3-7}
     & & \multicolumn{1}{c}{dev} & \multicolumn{1}{c}{dev2} & \multicolumn{1}{c|}{test} & \multicolumn{1}{c}{devtest} & \multicolumn{1}{c}{evltest} \\ 
    \hline\hline
    Transformer & 33.3 & 48.70 & 48.56 & 47.95 & 18.53 & 18.61 \\
    \text{Conformer-small} & 30.9 & 49.96 & 50.80 & 50.23 & 19.51 & 19.19 \\
    Conformer & 43.5 & \textbf{51.14} & \textbf{51.59} & \textbf{51.03} & \textbf{19.97} & \textbf{20.44} \\
    \bottomrule[1.5pt]
    \end{tabular}
}
\end{table}

\vspace{-0.2cm}
\section{SPEECH SEPARATION EXPERIMENTS}
\label{sec:ss-exp}
For the SS task, we compare our Conformer model with Transformer and bidirectional long short-term memory (BLSTM) on \href{https://www.merl.com/demos/deep-clustering/}{WSJ0-2mix} corpus. Both models are trained with uPIT~\cite{kolbaek2017multitalker} based on Phase Sensitive Masks (PSM) and ReLU activation function. The input features are 129-dimensional short-time Fourier transform (STFT) magnitude spectra computed with a sampling frequency of 8 kHZ, a frame size of 32 ms, and a 16 ms frame shift. The BLSTM-uPIT model has 3 BLSTM layers ($d=896$), while the Transformer-uPIT and Conformer-uPIT model consist of 3 blocks ($d^{\text{ff}}=896, d^{\text{att}}=1024, H=8$).

Table~\ref{table:ss-results} summaries the Signal-to-Distortion Ratio (SDR)~\cite{vincent2006performance} results of different models on the WSJ0-2mix sets, the current benchmark dataset to validate monaural speech separation. The results show that our Conformer-uPIT model gets competitive results compared with the BLSTM-uPIT model and achieves a significant improvement over the Transformer-uPIT model.

\begin{table}[tbp]
\vspace{-0.2cm}
\centering
\caption{Signal-to-Distortion Ratio (SDR) results of different models on the WSJ0-2mix cv/tt sets using uPIT.}
\vspace{0.2cm}
\label{table:ss-results}
\resizebox{0.6\linewidth}{!}{
    \begin{tabular}{c|c|c} 
    \toprule[1.5pt]
    \multicolumn{1}{c}{Model} & \multicolumn{1}{c}{cv} & tt \\ 
    \hline\hline
    BLSTM-uPIT~\cite{kolbaek2017multitalker} & 9.5 & 9.5 \\
    \midrule
    BLSTM-uPIT (ours) & 10.4 & \textbf{10.3} \\
    Transformer-uPIT & 9.3 & 8.8 \\
    Conformer-uPIT & \textbf{10.5} & 10.2 \\
    \bottomrule[1.5pt]
	\end{tabular}
}
\vspace{-0.2cm}
\end{table}

\vspace{-0.2cm}
\section{TTS EXPERIMENTS}
\vspace{-0.1cm}
\label{sec:tts-exp}

\label{subsec:tts-setup}
In TTS experiments, we evaluate with three common corpora: \href{https://keithito.com/LJ-Speech-Dataset/}{LJSpeech} (22.05 kHz, English, 24 hours), \href{https://sites.google.com/site/shinnosuketakamichi/publication/jsut}{JSUT} (24 kHz, Japanese, 10 hours), and \href{https://www.data-baker.com/open_source.html}{CSMSC} (24 kHz, Mandarin, 12 hours), all of which consist of single female speaker speech.
We compare the Conformer-based non-autoregressive models with Transformer-based models, including Transformer-TTS~\cite{li2019neural}, FastSpeech (FS)~\cite{ren2019fastspeech}, and FastSpeech2 (FS2)~\cite{ren2020fastspeech}.
We used $d^\mathrm{att}=368$, $d^\mathrm{ff}=1536$, and $H=2$.
The number of encoder and decoder blocks was set to six for FS, and four for FS2.
For FS2, instead of the quantized pitch and energy prediction, we use the token-average pitch and energy prediction introduced in \cite{lancucki2020fastpitch} to avoid overfitting.
As for the nerual vocoder, we use an open-source implementation\footnote{\url{https://github.com/kan-bayashi/ParallelWaveGAN}} of Parallel WaveGAN~\cite{yamamoto2020parallel}. 
All of the models are implemented with ESPnet2 except for the neural vocoder.
Training configurations, generated samples, and pre-trained models are available on Github\footref{foot:espnet}.

Table~\ref{table:tts-results} shows mel-cepstral distortion (MCD), which was calculated with 0-34 order mel-cepstrum and dynamic time warping (DTW) to match the length between the groundtruth and the prediction.
The result demonstrates that the Conformer-based models always bring consistent improvement for all corpora, achieving the best performance among the compared models.

\begin{table}[htbp]
\centering
\caption{MCD [dB] results on various open source TTS corpora. All models used a phoneme sequence as the input.}
\vspace{0.2cm}
\label{table:tts-results}
\resizebox{0.8\linewidth}{!}{
    \begin{tabular}{c|c|c|c} 
    \toprule[1.5pt]
    \multicolumn{1}{c}{Model} & \multicolumn{1}{c}{LJSpeech} & \multicolumn{1}{c}{JSUT} & CSMSC \\ 
    \hline\hline
    Transformer-TTS & 7.26 & 7.10 & 6.61 \\
    Transformer-FS & 6.91 & 6.75 & 6.27 \\
    Conformer-FS & 6.88 & 6.69 & \textbf{6.21} \\
    Transformer-FS2 & 6.85 & 6.69 & 6.30 \\
    Conformer-FS2 & \textbf{6.79} & \textbf{6.56} & 6.25 \\
    \bottomrule[1.5pt]
    \end{tabular}
}
\vspace{-0.3cm}
\end{table}

\vspace{-0.2cm}
\section{CONCLUSION}
\label{sec:conlusion}

We conducted comparative studies of the Conformer model in various speech applications with a large number of publicly available corpora. Specifically, the experiments were conducted on 25 ASR corpora (17 common sets + 8 low-resource sets), 1 ST corpus, 1 SS corpus, and 3 TTS corpora. From the experiments, our Conformer-based models achieved significant improvements in many ASR, ST and TTS tasks and competitive results in SS tasks. We believe that the various benchmark results, reproducible recipes, well-trained models and training tips described in this paper will accelerate the Conformer research on speech applications.
Our aim for this activity is to fill out the gap between high-resource research environments in big players and those in the academia or small-scale research groups by providing these up-to-date research environments.


\ninept
\bibliographystyle{IEEEbib}
\bibliography{strings,refs}

\end{document}